\documentclass[aps,prl,twocolumn,showpacs,showkeys,nofootinbib,superscriptaddress]{revtex4-1}

\usepackage[utf8]{inputenc}
\usepackage{graphicx}
\usepackage{amsfonts}
\usepackage{amssymb}
\usepackage{amsmath}
\usepackage[mathscr]{eucal}
\usepackage{setspace}
\usepackage{booktabs}
\usepackage[shellescape,latex]{gmp}
\usempxpackage{amssymb}

\def\sqr#1#2{{\vcenter{\hrule height.#2pt
   \hbox{\vrule width.#2pt height#1pt \kern#1pt
      \vrule width.#2pt}
   \hrule height.#2pt}}}

\def\bsqr#1#2{{\vrule width #1pt height#2pt}}
\def\bsquare{{\mathchoice\bsqr66\bsqr66\bsqr33\bsqr33}}
\def\badbreak{\penalty1000}

\newcommand{\cP}{{\cal P}}                    % cal-P
\newcommand{\cS}{{\cal S}}                    % cal-S
\newcommand{\cL}{{\cal L}}                     % cal-L
\newcommand{\msL}{{\mathscr L}}          % different cal-L
\def\fir{{\scriptscriptstyle{\text{\rm IR}}}}             % IR subscript 
\def\lm0{{\lambda_0}}                                     % IR scale 1
\def\nrN{N}                                               % number of objects
\def\cf{\mathfrak{n}}                                     % counting function
\def\cfu{\cf_\star}                                       % universal counting function
\def\efN{\mathscr{N}}                                     % effective number of objects
\def\efNm{\efN_\star}                                     % minimal effective number of objects
\def\pN{\mathscr{N}_p}                                    % participation number
\def\Obj{O}                                                     % set of objects
\def\w{c}                                                         % counting weight
\def\v{b}                                                            % counting weight
\makeatletter
\newcommand*{\smallrel}[2][.8]{%
  \mathrel{\mathpalette{\smallrel@{#1}}{#2}}%
}
\newcommand*{\smallrel@}[3]{%
  \sbox0{$#2\vcenter{}$}%
  \dimen@=\ht0 %
  \raise\dimen@\hbox{%
    \scalebox{#1}{%
      \raise-\dimen@\hbox{$#2#3\m@th$}%
    }%
  }%
}
\makeatother

\usepackage{amsmath} % wonderful math package
\usepackage{dsfont}  % double strike font
\usepackage{bm}      % bold math

\def\beq{\begin{equation}}
\def\eeq{\end{equation}}
\def\beqs#1\eeqs{\beq\begin{split} #1 \end{split}\eeq}

\long\def\comment#1{}

\def\be{\begin{equation}}
\def\ee{\end{equation}}

\def\bc{\begin{center}}
\def\ec{\end{center}}

\usepackage{microtype}
\usepackage[dvipsnames]{xcolor}
\usepackage[colorlinks=true,backref=false, linktocpage=true,
citecolor=PineGreen,urlcolor=PineGreen,linkcolor=PineGreen,pdfpagemode=UseOutlines]{hyperref}

\hypersetup{%
  bookmarksnumbered=true,
  pdftitle = {},
  pdfsubject = {},
  pdfauthor = {},
  pdfkeywords = {}
}

\begin{document}

\title{Low-Dimensional Life of Critical Anderson Electron}

\author{Ivan Horv\'{a}th}
\email{ihorv2@g.uky.edu}
\affiliation{Nuclear Physics Institute CAS, 25068 \v{R}e\v{z} (Prague), Czech Republic}
\affiliation{University of Kentucky, Lexington, KY 40506, USA}

\author{Peter Marko\v{s}}
\email{peter.markos@fmph.uniba.sk}
\affiliation{Dept. of Experimental Physics, Faculty of Mathematics, 
Physics and Informatics, Comenius University in Bratislava, Mlynsk\'a Dolina 2, 
842 28 Bratislava, Slovakia}

\date{Dec 8, 2022}

\begin{abstract}

We show that critical Anderson electron in 3 dimensions is present in its spatial
effective support, which was recently determined to be a region of 
fractal dimension $\approx \! 8/3$, with probability 
$1$ in infinite volume. Hence, its physics is fully confined to space of this lower 
dimension. Stated differently, effective description of space occupied by critical 
Anderson electron becomes a full description in infinite volume. We then show 
that it is a general feature of the effective counting dimension underlying these 
concepts, that its subnominal value implies an exact description by effective support.
 
%\smallskip


\end{abstract}

\maketitle

Disorder-induced localization of electrons, namely the Anderson 
transition~\cite{Anderson:1958a}, attracts the attention of theoretical 
physicists for over six decades. Its scaling theory~\cite{Abrahams:1979a} is based 
on the system size and the disorder dependence of electron 
conductance~\cite{Edwards_1972, Licciardello_1975, Slevin_2001}, and 
enables a quantitative description of the 
phenomenon~\cite{MacKinnon:1981a, Shklovskii:1993a}, including 
the calculation of critical exponents~\cite{Wegner:1989a}. However, 
the absence of self-averaging in the localized state~\cite{Anderson:1980a} 
suggests that faithful description of the transition requires information 
on statistical properties of key quantities, such as 
conductance~\cite{Markos:1993a,Markos:1999a}, and of electronic eigenstates 
themselves~\cite{Mildenberger:2002a,Vasquez:2008a,Rodriguez:2008a}.
Here we study critical wave functions that have been of interest, 
among other things, due to their intriguing geometric structure
\cite{Aoki_1983, Soukulis:1984a, Castellani:1986, Evangelou_1990, 
Schreiber:1991a,Janssen:1994}. 
Being more complex than that of a scale invariant fractal, this structure is commonly 
termed as multifractal. Its chief manifestation is the presence of non-trivial 
dimensional features.

Recently, a new type of dimension based on effective 
counting~\cite{Horvath:2018aap,Horvath:2018xgx,Horvath:2019qeo} has 
been proposed~\cite{Alexandru:2021pap, Horvath:2022ewv}. This effective 
counting dimension is analogous to Minkowski box-counting dimension of 
fixed sets~\cite{falconer2014fractal}. To explain the connection in the present 
setting, consider first the nominal dimension $D_\fir$ of spatial lattice $\msL$, 
which conveys the increase of lattice volume $\nrN$ (number of lattice sites) 
with linear size $L$, namely $\nrN[\msL(L)] \!\propto\! L^{D_\fir}$ 
for $L \to \infty$. Hence, $D_\fir$ is based on the same scaling as the infrared 
version of Minkowski dimension in a special case when all existing ``boxes" 
are counted. Next, consider a subset $\cS(L) \subset \msL(L)$ containing 
$\nrN[\cS(L)]$ points selected by some rule at each $L$. The Minkowski 
dimension $d_\fir$ of the infrared target $\cS$ defined by $\cS(L)$ then 
expresses the asymptotic growth of $\nrN$ with $L$, namely 
\begin{equation}
   \nrN[\cS(L)] \propto L^{d_\fir[\cS]}
   \quad\, \text{for} \quad\,
   L \to \infty
   \label{eq:005}   
\end{equation}
In the relations above, $\nrN \!=\! \nrN[\ldots]$ is treated as a function on sets,
namely as the ordinary counting measure.

However, if we wish to characterize wave functions describing quantum particles
by such spatial dimension, we encounter a conceptual problem. Indeed, wave 
function informs on probabilities for a particle to reside in any region of space
rather than on a unique subregion $\cS$ associated with the particle. Is it then 
even meaningful to talk about a volume-based characteristic analogous 
to Minkowski dimension?
 
Surprising answer to this question is yes~\cite{Horvath:2022ewv}. 
The schematic logic leading to the resulting prescription is as follows. Consider 
a probability vector $P=(p_1, p_2, \ldots, p_\nrN)$ where $p_i = \psi^+\psi(r_i)$ 
is the probability of the particle in state $\psi$ to be at lattice site $r_i$. Assume 
further that we can count lattice points $r_i$ effectively, based on their 
relevance $p_i$. In other words, that we have a function $\efN \!=\! \efN[P]$
at our disposal, returning effective count of entries in $P$. This could be used
to define the effective spatial support $\cS[P]$ of a particle as a collection 
of $\efN[P]$ points $r_i$ with highest probabilities. If the effective number
function $\efN$ is additive, then $\cS[P]$ respects relevant measure 
considerations~\cite{Horvath:2022ewv}. For IR-regularized wave functions 
$\psi(L)$, the scaling of $\efN[P(L)]$ with $L$ would then represent the IR 
dimension $d_\fir$ of the target $\psi$ in this probabilistic setting. The key 
result of~\cite{Horvath:2022ewv} is that all schemes $\efN$ producing 
valid supports $\cS[P]$ lead to the same $d_\fir$, which uniquely extends 
Minkowski dimension into the probabilistic domain. Value of $d_\fir$ 
can be extracted using the intrinsic (minimal) effective scheme $\efNm$ 
namely~\cite{Horvath:2018aap}
\begin{equation}
      \efNm[P]  \,=\, \sum_{i=1}^\nrN \cfu( \nrN p_i)   \quad,\quad
      \cfu(\w)  \; = \;   \min\, \{ \w, 1 \}    \;
      \label{eq:025}         
\end{equation}
More explicitly, the following analog of \eqref{eq:005}
\begin{equation}
   \efNm[P(L)]  \propto  L^{d_\fir[\psi]}   
   \quad\, \text{for} \quad\,
   L \to \infty
   \label{eq:045}   
\end{equation}
defines $d_\fir$ of the target wave function $\psi$.

We emphasize that the concept of $d_\fir[\psi]$ is recent
\cite{Horvath:2018aap,Horvath:2018xgx,Horvath:2019qeo,
Alexandru:2021pap, Horvath:2022ewv} and has so far been applied
to Dirac modes in quantum chromodynamics~\cite{Alexandru:2021pap}
and states at 3d Anderson transitions~\cite{Horvath:2021zjk}. Despite 
the extensive literature on the spatial structure in the latter context (e.g. 
\cite{Mildenberger:2002a, Vasquez:2008a, Rodriguez:2008a, Aoki_1983, 
Soukulis:1984a, Evangelou_1990, Schreiber:1991a, Janssen:1994,
Ujfalusi:2015a, Evers_2008,Markos:2006A}), information conveyed by 
$d_\fir$ is different and complementary to existing results due to additivity 
of effective counting. 

Recent calculation of $d_\fir$ for critical wave functions at 3d Anderson 
transitions produced unexpected results~\cite{Horvath:2021zjk}. 
To explain them, consider the 3d Anderson model~\cite{Anderson:1958a} 
on cubic lattice ($D_\fir \!=\!3$) with periodic boundary conditions, 
and the Hamiltonian (orthogonal class)
\begin{equation}
     {\cal H} \,=\, \sum_r \epsilon_r \, c^\dag_r  \, c_r 
     \,+\, \sum_{r,j}  c^\dag_r \, c_{r-e_j} + h.c.
     \label{eq:065}
\end{equation}
Here $r\!=\!(x_1,x_2,x_3)$ labels lattice sites, $e_j$ ($j\!=\!1,2,3$)~unit 
lattice vectors, $\epsilon_r \!\in\! [-W/2,+W/2]$ uniformly distributed random 
potentials, and $c_r$ the 1-component electron operators.
Definition of $d_\fir \!=\! d_\fir(E,W)$ involves averaging over states 
in the vicinity of energy $E$ and over disorder $\{ \epsilon_r \}$,  
thus entailing $\efNm \!\rightarrow\! \langle \,\efNm \rangle$ in 
Eq.~\eqref{eq:045}. At the critical point $E\!=\!0$, 
$W \!=\! W_c\!=\!16.543(2)$ \cite{Slevin_2018}, the calculation in 
Ref.~\cite{Horvath:2021zjk} yielded the dimension $d_\fir \!\approx\! 8/3$.  
In fact, the computed values in orthogonal, unitary, symplectic and chiral
(AIII) classes all turned out to be well within two parts per mill of 8/3, while 
being pairwise equal within errors. These findings led 
to the proposition that the critical exponent $d_\fir$ is super-universal in 
3d Anderson transitions~\cite{Horvath:2021zjk}.

Here we inquire about the probability $\cP$ that 3d critical Anderson electron 
is present in its spatial support, which conveys the extent to which is its 
physics tied to this low-dimensional ($d_\fir \!<\! D_\fir$) space. Formal
definition of $\cP$ relies on that of effective support~\cite{Horvath:2022ewv}. 
The latter starts by ordering lattice sites in $\msL =\{r_1,r_2, \ldots ,r_\nrN\}$ 
by their relevance, so that $p_1 \ge p_2 \ge \ldots \ge p_\nrN$ in $P$. 
To realize supports involving both integer and non-integer number of sites 
($\efN$ is real-valued), we represent them by generalized collections 
$\{r_1, \ldots ,r_J : f\}$, where $0 \!<\! f \!\le\! 1$ specifies 
the fraction of $r_J$ included. The effective support of $P$ 
(equivalently of $\psi$) on $\msL$ in counting scheme $\efN$ is 
\footnote{While the distinction between full and fractional
inclusion of $r_J$ doesn't matter for most asymptotic ($\nrN \!\to\! \infty$) 
considerations, it affects the accuracy of extrapolations when $d_\fir$ 
is much smaller than $D_\fir$. Fractional approach keeps additivity relations
exact~\cite{Horvath:2022ewv}.}
\begin{equation}
   \cS[P,\efN\,] \,=\, \{\,r_1, \ldots ,r_J \,:\, \efN[P] + 1 \!-\! J \,\} 
   \label{eq:185}   
\end{equation}
where $J \!=\! \text{\tt ceil} ( \efN[P] )$, and {\tt ceil} is the ceiling function.
Note that $J \!\le\! \nrN$. The probability associated with $\cS$ is
\begin{equation}
   \cP[P,\efN] \,=\, \sum_{j=1}^{J-1}\, p_j   \,+\,  f \, p_J
   \label{eq:085}   
\end{equation}
where $J\!=\!J[P,\efN]$ and $f\!=\!f[P,\efN]$ are specified by Eq.~\eqref{eq:185}.

\begin{figure}[t]
   \includegraphics[width=0.395\textwidth]{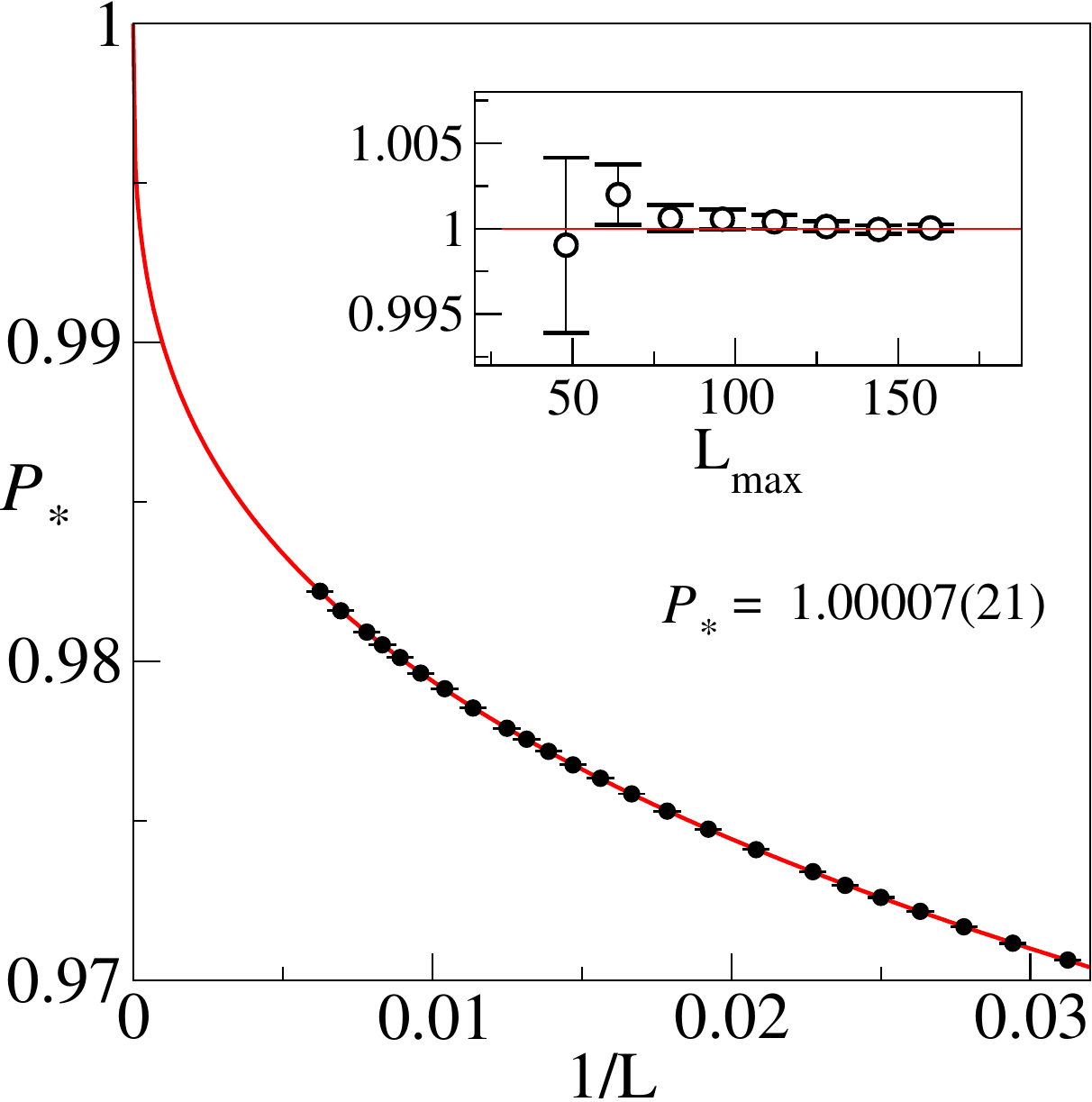}
   \vskip -0.05in   
   \caption{Probability $\cP_\star$ that critical Anderson electron is found 
   in its minimal effective support $\cS_\star$. Fit to $\cP_\star + b/L^\beta$ 
   for all data ($L$ in range $[32,160]$) and the resulting $\cP_\star$ are 
   shown. Fits in range $[32,L_{\text{max}}]$ lead to $\cP_\star$ values shown 
   in the inset.}
   \label{fig:pstar}
   \vskip -0.17in
\end{figure}

%We now determine this probability for critical Anderson electron 
%(Hamiltonian \eqref{eq:065}) and its effective support $\cS_\star$ defined 
%by the intrinsic scheme $\efNm$ via Eqs.~\eqref{eq:185} and \eqref{eq:085}. 
%More precisely, we numerically compute the critical value
We now determine this probability for critical Anderson electron 
(Hamiltonian \eqref{eq:065}) in the intrinsic scheme $\efNm$, i.e.
$\cS[P,\efN] \!\to\! \cS_\star[P] \!\equiv\! \cS[P,\efNm]$ and
$\cP[P,\efN] \!\to\! \cP_\star[P]$ in  Eqs.~\eqref{eq:185}, \eqref{eq:085}. 
More precisely, we compute the critical~value
\begin{equation}
    \cP_\star \equiv \lim_{L \to \infty}  \, \langle\, \cP_\star[P(L)] \,\rangle
    \label{eq:095}        
\end{equation}
where averaging has the same meaning as in the case of critical dimension 
$d_\fir$.  In what follows, $L$ is dimensionless i.e. expressed in units of 
lattice spacing. We study 24 finite systems with sizes in the range 
$32 \!\le\! L \!\le\! 160$, each represented by at least 40k realizations of 
disorder. Electron eigenfunctions were computed using 
the JADAMILU library~\cite{jadamilu_2007}. Two levels closest to 
$E\!=\!0$ for each disorder were included in the~analysis.

Computed probabilities are shown in Fig.~\ref{fig:pstar} as a function of 
$1/L$. The asymptotic $1/L \!\to\! 0$ value $\cP_\star$ is extracted by 
fitting to $\cP_\star + b/L^\beta$ involving a general power correction. 
Since data at different $L$ are uncorrelated, we use the standard 
minimum $\chi^2$ procedure. The master fit using all data leads 
to $\cP_\star \!=\! 1.00007(21)$. To see how this emerges as the size
of the system increases, we show in the inset of Fig.~\ref{fig:pstar} 
the values of $\cP_\star$ obtained from fits in intervals $[32,L_{\text{max}}]$ 
where $L_{\text{max}} \!=\! 48,64,80, \ldots,160$. All fits were stable and 
had good $\chi^2$/DOF ($0.8 \!<\! \chi^2/\text{DOF} \!<\! 1.1$). Given 
the accuracy and robustness of these results, and that 
$\lim_{L_{\text{max}} \to \infty} \cP_\star(L_{\text{max}})\!=\! \cP_\star$, 
we conclude that $\cP_\star \!=\!1$.

Few comments are in order. 
(i) We found that the association (expressed by $\cP_\star$) of critical 
Anderson electron with its spatial support $\cS_\star$ increases with 
system size $L$, although the volume of $\cS_\star$ shrinks relative to 
the volume of the lattice. Indeed, the ratio 
$\nrN[\cS_\star(L)]/\nrN[\cL(L)]$~approaches zero 
roughly as $L^{-1/3}$ since $D_\fir \!=\! 3$ and $d_\fir \!\approx\! 8/3$.
(ii) Moreover, electron becomes fully confined to $\cS_\star$ in 
$L \!\to\! \infty$ limit ($\cP_\star \!\to\! 1$). This space of lower dimension 
thus turns into a complete arena for its physics.
(iii) Given that the counting scheme $\efNm$ is minimal~\cite{Horvath:2018aap},
the associated effective support $\cS_\star$ is also 
minimal~\cite{Horvath:2022ewv}, together with its $\cP_\star$. 
More precisely, we have
\begin{equation}
    \cP_\star[P] = \cP[P,\efNm] \,\le\, \cP[P, \efN \,]
    \label{eq:105}        
\end{equation}
for all distributions $P$ on a finite lattice, and all $\efN$ consistently defining 
effective support~\cite{Horvath:2022ewv}. Hence, $\cP_\star \!=\! 1$
implies $\cP \!=\! 1$ for all valid $\efN$. In other words, similarly to 
$d_\fir \!\approx\! 8/3$,  the property $\cP \!=\!1$ is well-defined.

\begin{figure}[t]
   \includegraphics[width=0.412\textwidth]{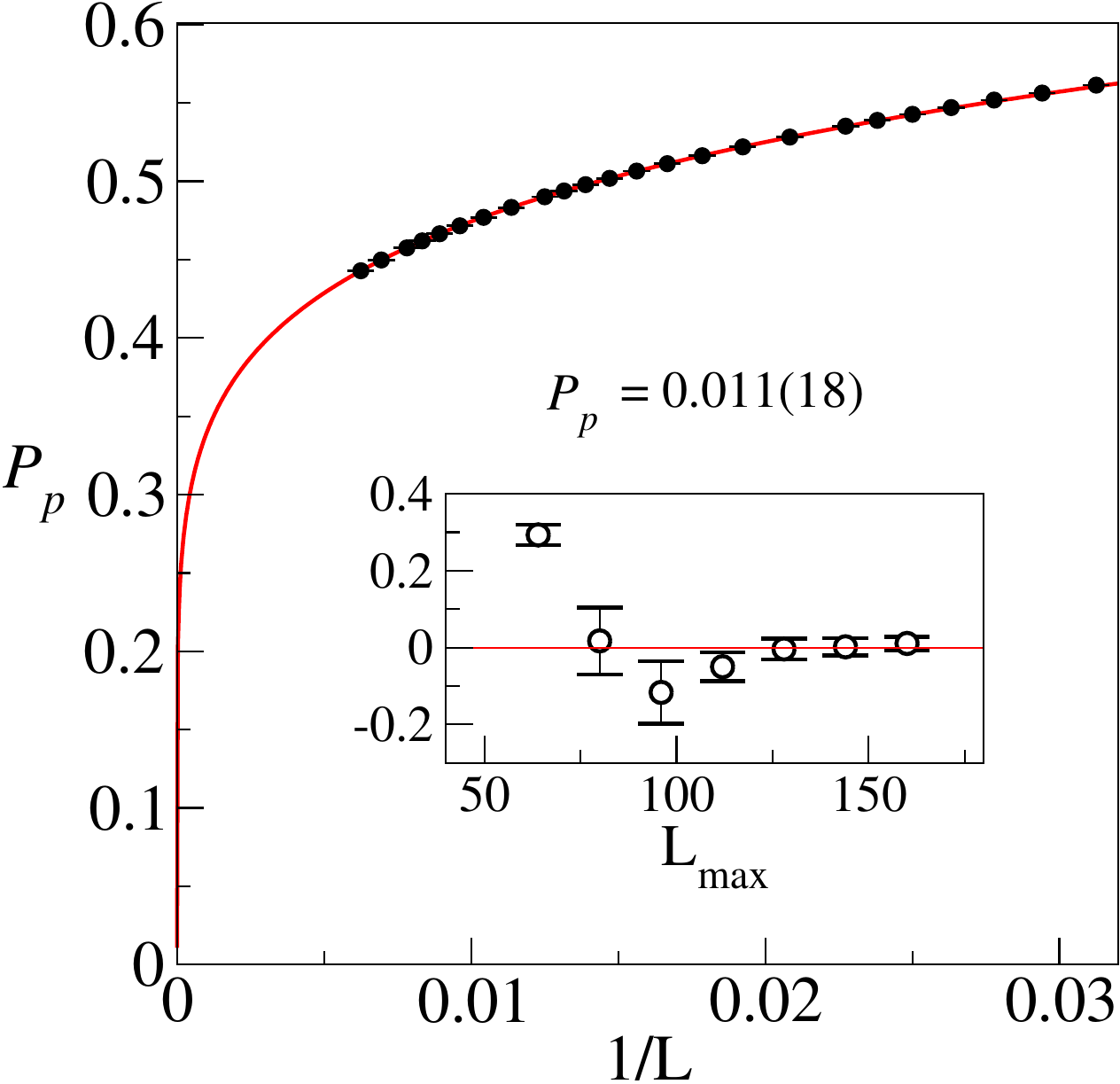}
   \vskip -0.05in   
   \caption{Probability $\cP_p$ that critical Anderson electron is found
   in region $\cS_p$ specified by the participation number $\efN_p$. Fit to
   $\cP_p + b/L^\beta$ for $L$ in range $[48,160]$ and the resulting
   $\cP_p$ are shown. Fits in range $[48,L_{\text{max}}]$ produce $\cP_p$ 
   values in~the~inset.}
   \label{fig:ppn}
   \vskip -0.16in
\end{figure}

We now show that $\cP_\star \!=\! 1$ follows from $d_\fir \!< D_\fir$. 
In fact, the corresponding feature is built into the effective dimension 
already at the pre-metric level: aspects involving distance, such as infrared
and ultraviolet, are not crucial to it. We thus give the argument 
in the most general setting where collections $O=\{o_1,o_2, \ldots o_\nrN\}$ 
contain arbitrary objects ordered by $P$, with no additional structure 
assumed. The notions of effective support $\cS[P,\efN]$ and its probability 
$\cP[P,\efN]$ (Eqs.~\eqref{eq:185} and \eqref{eq:085}) generalize trivially, 
with all observations made so far still holding. Effective counting dimension
in this setting expresses the scaling of $\efNm$ with $\nrN$ 
(see \cite{Horvath:2022ewv} and below).

Our reasoning relies on the relationship between $\efNm$ and $\cP_\star$ 
that follows directly from the defining relations~\eqref{eq:025}, 
\eqref{eq:185} and \eqref{eq:085}. In particular, splitting $\efNm$ into 
a contribution $\efNm^\cS$ from objects in support $\cS_\star$ and 
the rest yields
\begin{equation}
    \efNm[P\,] \,=\, \efNm^\cS[P\,]  + \nrN \bigl( 1 - \cP_\star[P\,]   \, \bigr )  
    \label{eq:115}        
\end{equation}
for all $P \!=\! (p_1,p_2, \ldots , p_\nrN)$. Indeed, note first that $\cS_\star$
contributes $\efNm^\cS \!=\! \sum_{i=1}^{J-1}\cfu(\nrN p_i) + f\, \cfu(\nrN p_J)$
and the rest is $(1-f)\, \cfu(\nrN p_J) + \sum_{i=J+1}^\nrN \cfu(\nrN p_i)$.
Since $\cfu(\nrN p) \!=\! \nrN p$ for $p \le 1/\nrN$ and $p_i \le 1/\nrN$
for $i\!=\!J,J+1,\ldots,\nrN$,
we obtain the formula \eqref{eq:115}.

Effective counting dimension $\Delta$ is assigned to a sequence 
$\Obj_k$ of collections with increasing number $\nrN_k$ of objects
and probability vectors 
$P_k \!=\! (p_{\scriptscriptstyle{k,1}}, p_{\scriptscriptstyle{k,2}}, 
\ldots, p_{\scriptscriptstyle{k,\nrN_k}})$. The pair $\Obj_k$, $P_k$ 
``regularizes" the $k \!\to\! \infty$ target defined by the sequence. 
Dimension $\Delta$ of the target is specified by~\cite{Horvath:2022ewv}
\begin{equation}
   \efNm[P_k] \propto \nrN_k^{\,\Delta}
   \quad\text{for}\quad k \to \infty
   \label{eq:060}
\end{equation}
thus taking values in the range $0 \!\le\! \Delta \!\le\! 1$. The subdimensional 
case $\Delta \!<\! 1$ implies $\cP_\star \!=\! 1$. To see that,  
write Eq.~\eqref{eq:115} for $P_k$ in the form
\begin{equation}
    \frac{\efNm[P_k] - \efNm^\cS[P_k]}{\nrN_k}  \,=\, 1 - \cP_\star[P_k]  
    \label{eq:125}        
\end{equation}
If $\Delta \!<\! 1$ then lhs approaches zero at least as fast as 
$(1/\nrN_k)^{1-\Delta}$ for $k \!\to\! \infty$. 
Hence, $\lim_{k \to \infty} \cP_\star[P_k] \equiv \cP_\star = 1$.

When metric is in place and regularization removal $k \!\to\! \infty$ 
corresponds to infinite-volume (IR) limit $L_k \!\to\! \infty$, then 
$d_\fir$ of Eq.~\eqref{eq:045} and $\Delta \!\equiv\! \Delta_\fir$ of 
Eq.~\eqref{eq:060} are related by $d_\fir \!=\! \Delta_\fir D_\fir$.
Thus, $\Delta_\fir \!<\! 1$ is equivalent to $d_\fir \!<\! D_\fir$.
Hence, the latter implies $\cP_\star \!=\! 1$ for critical 
Anderson electrons, as claimed. 

\smallskip

\noindent We close with a few comments. 

\smallskip

\noindent
{\bf (a)} We showed that, in the subdimensional case, the minimal 
effective support $\cS_\star$ fully describes collection $\Obj$ of 
$\nrN$ objects when $\nrN \!\to\! \infty$. 
Consequently, $\cS$ based on counting schemes other than 
$\efNm$, and thus entail more voluminous descriptions, involve 
a redundancy:  any excess of probability in $\cS$ relative 
to $\cS_\star$ goes to zero in $\nrN \to \infty$ limit when 
$\Delta \!<\! 1$. This further underlines the unique standing of 
$\efNm$ and $\cS_\star$ in effective quantitative~analyses.

\noindent
{\bf (b)} We focused on the Anderson system with $\cS_\star$ describing 
the physical space occupied by the electron. The critical case is 
subdimensional ($d_\fir \!\approx\! 8/3$~\cite{Horvath:2021zjk})
and the above property implies that electron is fully confined to
the corresponding subvolume in $L \!\to\! \infty$ limit. This is confirmed
by our detailed numerical analysis. Since simulations of Anderson 
systems provide the full access to $\cS_\star$, all aspects of critical 
spatial geometry and their relationship to the underlying physics can in 
fact be investigated.

\noindent
{\bf (c)} The value $d_\fir \!\approx\! 8/3$ may be connected to 
anomalous scaling of critical diffusion length 
$\ell(t) \propto t^{1/3}$~\cite{Sheng:2006a}. (Possibly also to exponent 
$4/3$ characterizing percolation clusters~\cite{Alexander:1982a}
(see also~\cite{Aharony:1987a}).) This is plausible since anomalous 
diffusion is likely due to electron propagation being effectively restricted 
to wave function support on a sample. That, in turn, qualitatively relates 
to the associated diffusion properties. Such connection is also 
corroborated by the observed superuniversality of 
$d_\fir$~\cite{Horvath:2021zjk} since the diffusion exponent is also 
expected to be superuniversal. The present work fortifies this possible 
connection. Indeed, not only is $\cP_\star \!=\! 1$ super-universal in light 
of the general argument given here, we also expect it to be a necessary 
ingredient in any microscopic derivation of the anomalous diffusion 
exponent. 

\noindent
{\bf (d)} Electron states at Anderson transitions are frequently
characterized via the participation number of Bell and Dean,
$\pN[P]=1/\sum_{i=1}^{\nrN} p_i^2$~\cite{Bell:1970a}, 
and the associated dimension $d_p$ extracted from 
$\pN \!\propto\! L^{d_p}$. Applying this to our data we obtained 
$d_p \!=\! 1.305(2)$. Although not an effective counting scheme, 
$\pN$ satisfies all required axioms except 
additivity~\cite{Horvath:2018aap}. Its comparison to $\efNm$ thus 
tests the importance of preserving the measure property. Proceeding 
as in the case of effective counting schemes, we define the probability 
$\cP_p$ within the ``support" $\cS_p$ based on $\pN$. These probabilities 
exhibit larger fluctuations at criticality than $\cP_\star$ and are shown in 
Fig.~\ref{fig:ppn}. In a striking difference to the additive case, 
$\langle \,\cP_p[P(L)]\, \rangle$ decreases with increasing $L$ and 
may converge to zero in $L \!\to\! \infty$ limit. Hence, the relevance 
of $\cS_p$ as a descriptor of space in which the physics of Anderson 
critical electron takes place is rather limited. 

\noindent
{\bf (e)} Multifractal formalism
(see e.g.~\cite{falconer2014fractal, Halsey_Kadanoff_multif:1986}) 
is also frequently used in analyses of Anderson criticality. Conceived 
in the context of turbulence and strange attractors, it describes local 
singularities of associated complex measures in a statistical manner. 
The original UV form of multifractality has been adopted for Anderson 
criticality~\cite{Castellani:1986, Evangelou_1990, Schreiber:1991a, 
Janssen:1994} via replacement $a \to \lambda \!=\! 1/L$ 
(or $\ell/L$ with coarse-grained scale $\ell$), and scaling for 
$\lambda \to 0$. In physics terms, release of UV cutoff ($a \!\to\! 0$) 
is replaced by that of IR cutoff ($L \!\to\! \infty$). 
Very recently, Ref.~\cite{Burmistrov:2022} proposed the formula for
average $\efNm$ in terms of multifractal parameters obtained by 
the moment method. Its reliability and impact has been questioned 
in Ref.~\cite{Horvath:2022d}.

\begin{acknowledgments}
   P.M. was supported by Slovak Grant Agency VEGA, Project n. 1/0101/20.
   We thank Robert Mendris for useful comments.
\end{acknowledgments}

\bibliography{my-references}

\end{document}